\documentclass[balance,upint,subscriptcorrection,varvw,mathalfa=cal=boondoxo,spanish,french,vietnamese,russian,greek,pdf-a,colorlinks]{asmeconf}

\hypersetup{%
	pdfauthor={John H. Lienhard},
	pdftitle={ASME Conference Paper LaTeX Template},                  % <=== change to YOUR pdf file title
	pdfkeywords={ASME conference paper, LaTeX template, BibTeX style},% <=== change to YOUR pdf keywords
	pdfsubject = {Describes the asmeconf LaTeX template},			  % <=== change to YOUR subject
	pdflicenseurl={https://ctan.org/pkg/asmeconf},% may delete
}

\begin{document}

% Change these fields to the right content for your conference.
% You can comment these out if for some reason you don't want a header.
% Use title case (first letters capitalized), not all capitals

% \ConfName{Proceedings of the ASME 2023\linebreak International Mechanical Engineering Congress and Exposition}
% \ConfAcronym{IMECE2023}
% \ConfDate{October 29--November 2, 2023} % update 
% \ConfCity{New Orleans, LA} % update 
% \PaperNo{IMECE2023-XXXX}

% Units of measure (e.g., cm) and other specialty lowercase terms in the title should be 
%   enclosed in \NoCaseChange{...} to maintain lower case type
%   LaTeX will automatically set the rest of the title in all capital letters.

\title{Why Do Humans Twist Their Ankle: A Nonlinear Dynamical Stability Model For Lower Limb} % <=== replace with YOUR title
%\title{Place Title Here: Place Subtitle After Colon} 
 
%   Put author names into the order you want. Use the same order for affiliations.
%   \affil{#} tags the author's affiliation to the address in \SetAffiliation{#}.
%   No space between last name and \affil{#}, separate names with commas.
%
%	For a sole author or a single affiliation for all authors, {#} may be left empty, as \affil{} and \SetAffiliation{} (but not with [grid] option!)
%
%   \CorrespondingAuthor{email} follows that author's affiliation, no spaces.  
%   If multiple corresponding authors, put both email addresses in the same command and place after both authors.
%
%   \JointFirstAuthor, if applicable, follows the affiliation of the relevant authors, no spaces.

\SetAuthors{%
	Yue Guan\affil{1}\CorrespondingAuthor{yguan1@memphis.edu}
	}

\SetAffiliation{1}{University of Memphis, Memphis, TN }
%	Note: Luis and Maria are not real people.  Henry and Catherine have been dead for >450 years.

%	To switch from inline author names to gridded names, use the [grid] option.

\maketitle

%%% Use this footnote for tracking various versions of your draft. Change text to suit your own needs. 
%%% \date{..} calls the same command. 
\versionfootnote{Documentation for \texttt{asmeconf.cls}: Version~\versionno, \today.}% <=== Delete before final submission.

Ankle injuries, which make up approximately 40\% of all the lower extremity injuries, are among the most common types of sports injuries \cite{fernandez2007epidemiology}.
Although ankle injuries resulting from postural instability are frequently observed during high-speed and intense physical activities, most current research has been limited to static or quasi-static models of the lower limb or has focused solely on the ankle joint itself.
Several standard techniques are available to measure postural stability, which refers to an individual's ability to maintain an upright and stable posture even after experiencing unpredictable perturbations.
The static standing steadiness test assesses the subject's ability to maintain stillness \cite{riemann1999examination}, but it overlooks the role of gravity and kinematic mechanisms in falling.
Tests that assess dynamic postural stability, such as the Star Excursion Balance Test (SEBT) \cite{tahmasebi2015evaluation, sell2012examination}, are actually quasi-static tests that provide adequate safety for the subjects but differ significantly from actual instability scenarios.
In this study, in order to explain the kinetic mechanism underlying postural instability and ankle twists, we present a nonlinear dynamical model for the human lower limb that takes into account transient behavior and large rotations.

\begin{figure}[hbp]
\centering\includegraphics[width=1\linewidth]{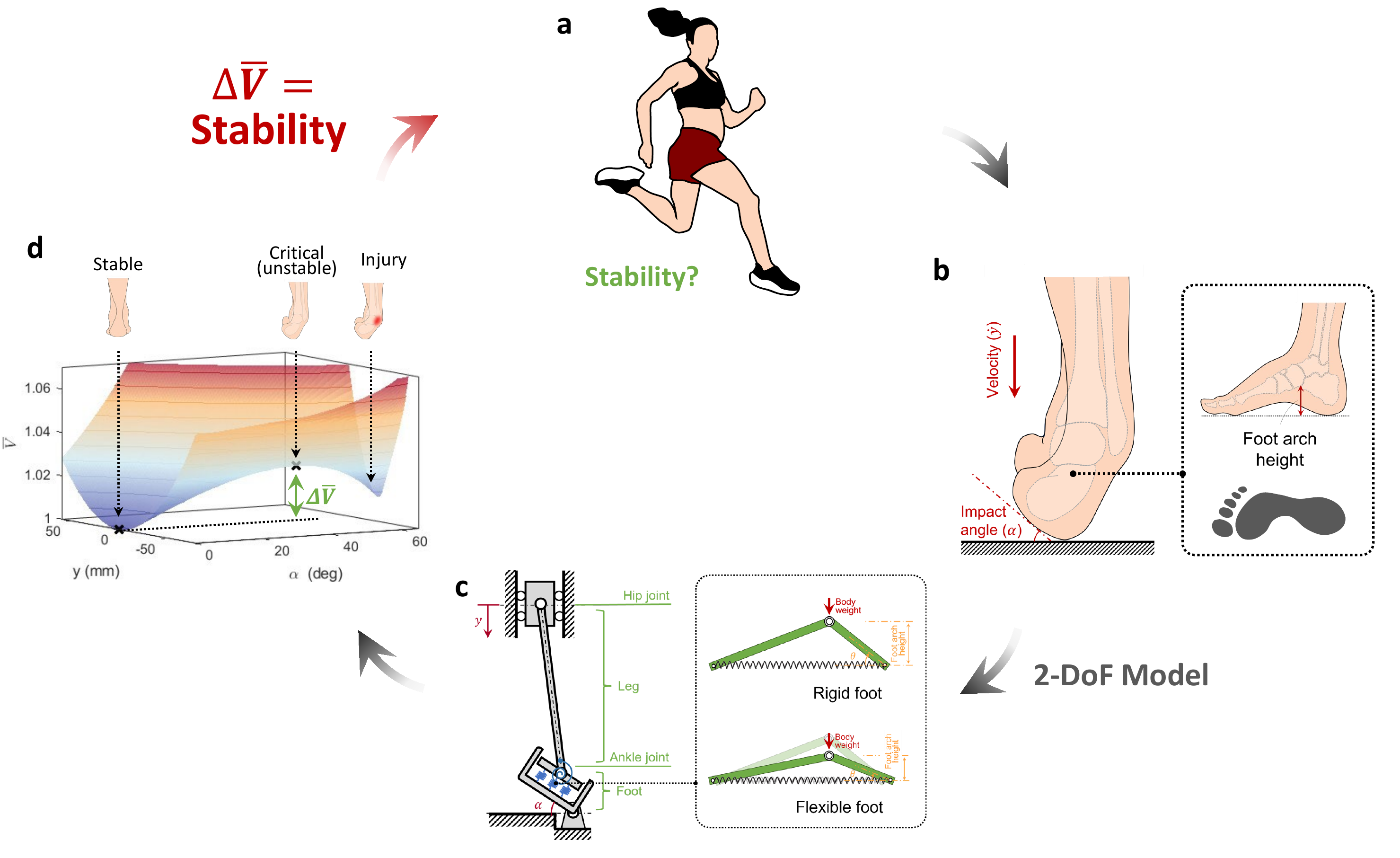}
\caption{A nonlinear dynamical model for human lower limb. a, the motion pattern under study. b, Critical parameters in dynamical analysis. c, The 2-DoF nonlinear dynamical model. d, The 2-DoF potential energy landscape and stability indicator $\Delta \overline{V}$.}\label{fig:Intro}
\end{figure}

Ankle injuries are most commonly seen in activities involving jumping and landing \cite{fernandez2007epidemiology}. 
Here we focus on this specific motion pattern and examine the postural stability of the lower limb under different landing velocities ($\dot{y}$) and initial inclination angles ($\alpha$).
We assume that instability occurs in the coronal (frontal) plane.
The landing of the lower limb on the ground is modeled as a mechanical system with two Degrees of Freedom (DoFs).
As shown in Fig.~\ref{fig:Intro}~c, the leg is represented as a rigid bar, and the foot is modeled as an elastic foundation that allows movement in one direction.
The foot stiffness is assumed to be determined by the foot's medial longitudinal arch (MLA), which is simulated as an arch-spring mechanism.
We simplify the hip joint as a kinematic pair that permits vertical displacement and rotation, but restricts horizontal movement.
The ankle joint is modeled as a turning pair with a specified rotational stiffness.
When the lower limb experiences external perturbations, the foot begins to rotate around its lateral edge.
The configuration of the lower limb mechanism can be completely defined by two DoFs: the vertical displacement ($y$) and the inclination angle ($\alpha$).
The upright, stable configuration occurs at $\left(y, \alpha \right) = \left(0, 0 \right)$.
Clearly, the mechanism remains stable if the parameter trajectory $\left(y (t), \alpha (t) \right)$ returns to $\left(0, 0 \right)$ after an impact, whereas the mechanism loses stability if $\left(y (t), \alpha (t) \right)$ becomes unbounded.
Thus, the ankle twist scenario is described as a transient stability problem for a 2 DoF mechanical system.

%Inset, The foot arch height and its contribution to the lower limb dynamics are considered. 
%Inset, arch-spring mechanisms for foot structures. 

The transient response of the 2 DoF system after an external impact is determined by its potential energy landscape (shown in Fig.~\ref{fig:Intro}~d).
The upright, stable configuration presents a local minimum at $(0, 0)$ on the potential energy terrain. 
Thus, under small perturbations, the transient trajectory oscillates around and eventually ends at this local minimum, indicating that the individual maintains postural stability after the impact.
However, when the external perturbation energy increases, the trajectory may surpass the energy hilltop and escape from the potential energy minimum, resulting in a loss of stability and causing the individual to fall down.
The index-1 saddle point (equilibrium point with only one unstable eigenvalue) on the potential energy landscape plays a critical role in determining whether this transition occurs, acting as a mountain pass that must be overcome.
The minimum impact energy required to lose stability is equal to the potential energy level of the index-1 saddle point.
We normalize this minimum impact energy by the potential energy at $(0, 0)$ and use the resulting normalized energy level $\Delta \overline{V}$ as an indicator of lower limb postural stability.
A larger $\Delta \overline{V}$ indicates that the lower limb system can withstand larger perturbation energy without losing stability.

\begin{figure}
\centering\includegraphics[width=0.98\linewidth]{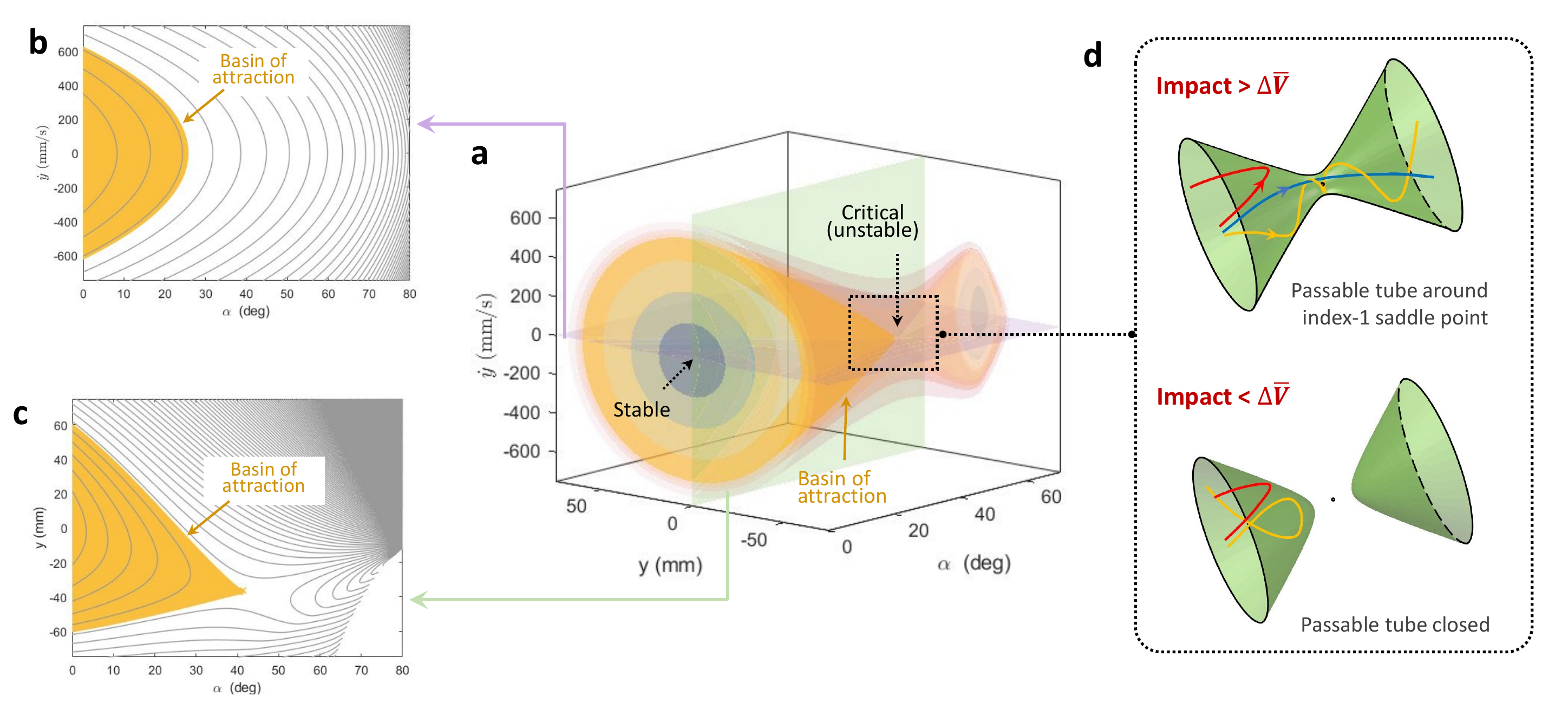}
\caption{Iso-potential shapes for human lower limb model and the basin of attraction for the upright posture. a, Iso-potential shapes in the 3D state space. b-c, Cross-sectional views of the basin of attraction (shaded) in falling velocity-inclination and falling displacement-inclination spaces respectively. d, The opening and closing of the passable tubes around index-1 saddle (critical) points.}\label{fig:PE}
\end{figure}
% Contour plots of the potential energy and 
% depends on the impact energy level.

Figure~\ref{fig:PE} shows a typical potential energy landscape for the lower limb system in the 3D state space.
In addition to the two displacement parameters ($y$ and $\alpha$), here the falling velocity ($\dot{y}$) is also taken into consideration.
Each iso-potential shape presents an outer boundary for the transient trajectories starting within it.
When the external impact energy is greater than the stability indicator $\Delta \overline{V}$, as shown in Fig.~\ref{fig:PE}~d, a hyperboloid-shaped passable tube opens around the index-1 saddle, which allows the transient trajectories to pass through and escape from the initial stable (upright) configuration.
On the contrary, when the external impact energy drops below $\Delta \overline{V}$, the passable tube closes, and no trajectory can escape from the initial local minimum.
Therefore, the iso-potential shape around the initial stable point at the level of $\Delta \overline{V}$ exhibits a basin of attraction (ocher region).
Any combination of initial falling velocity ($\dot{y} (0)$), initial falling displacement ($y (0)$), and initial inclination angle ($\alpha (0)$) that falls within the basin of attraction maintains stability after the transient behavior dies out.
Specifically, the cross-section views of the basin of attraction in the falling velocity-inclination ($\dot{y}-\alpha$) and falling displacement-inclination ($y-\alpha$) spaces are presented in Fig.~\ref{fig:PE}~b and c, respectively.
In practice, we assume the initial falling displacement $y(0)$ is zero.
Thus, the size of the 2D basin of attraction in Fig.~\ref{fig:PE}~b indicates the stability of the lower limb system.
According to the previous analysis, the basin of attraction size is dominated by the normalized energy threshold $\Delta \overline{V}$.
The analysis of the basin of attraction leads us to conclude that $\Delta \overline{V}$ is still the most suitable indicator of stability for the lower limb system.

\begin{figure}
\centering\includegraphics[width=0.8\linewidth]{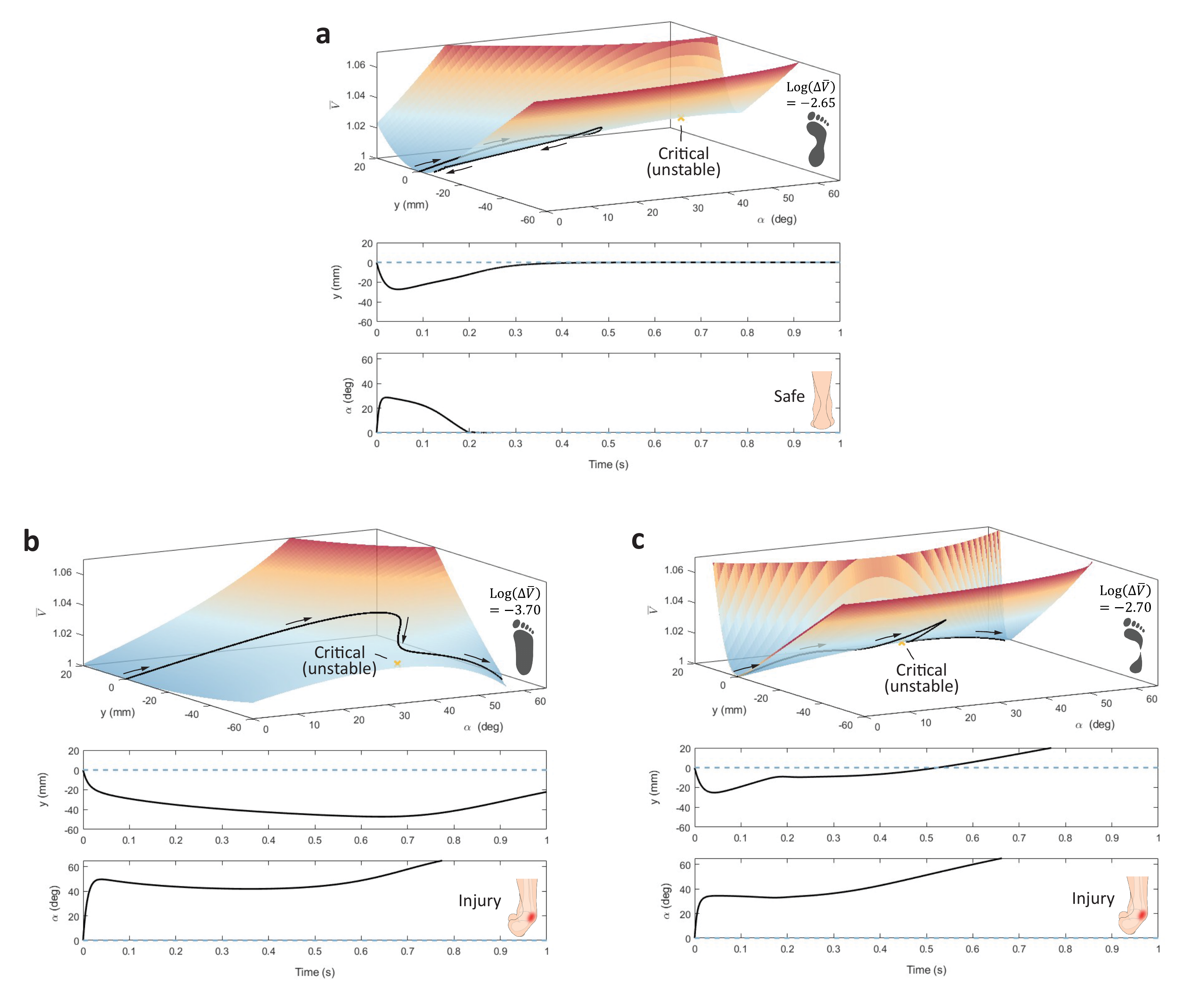}
\caption{A comparison of potential energy terrains and transient trajectories with identical initial conditions for normal foot (a), flat foot (b), and high arch foot (c) respectively.}\label{fig:Trans}
\end{figure}

Specifically, we use this stability indicator $\Delta \overline{V}$ to evaluate the postural stability of individuals with normal foot, flat foot, and high arch foot, respectively. 
As shown in Fig.\ref{fig:Trans}, the normal foot with a foot arch height ($h$) of $25mm$ presents the highest stability indicator $\Delta \overline{V}$, while both the flat foot ($h=5mm$) and high arch foot ($h=38mm$) are less stable than the normal foot.
In a trial test, all three lower limb systems with flat, normal, and high arch feet are subjected to an identical external impact.
Starting with the same initial condition, the trajectory of the normal foot system returns to the initial stable point after a transient oscillation, while both flat foot and high arch foot systems lose stability and their trajectories become unbounded.
The numerical example verifies the effectiveness of the stability indicator $\Delta \overline{V}$ and demonstrates that individuals with a normal foot have better postural stability compared to those with flat and high arch foot, and are less likely to experience ankle injuries in the jumping and landing scenario.

% \begin{figure}
% \centering\includegraphics[width=0.98\linewidth]{Stability-Arch-3.pdf}
% \caption{Lower limb morpho-stability relations. a, Lower limb stability $\lambda_s$ in the foot arch stiffness-height space. b-c, Foot arch angle ($\theta$) -lower limb stability ($\lambda_s$) relation for rigid and flexible foot mechanisms respectively. d-f, Basins of attraction for flat foot, normal foot, and high arch foot respectively under rigid foot mechanism. g-i, Basins of attraction for flat foot, normal foot, and high arch foot respectively under flexible foot mechanism.}\label{fig:Stability}
% \end{figure}

%\nocite{*}%% <=== Delete this line unless you want to typeset the entire contents of your .bib file!

\bibliographystyle{asmeconf}  %% .bst file following ASME conference format. Do not change.
\bibliography{asmeconf-guan}%% <=== change this to name of your bib file

\begin{thebibliography}{1}
\newcommand{\enquote}[1]{``#1''}
\providecommand{\url}[1]{\texttt{#1}}
\providecommand{\urlprefix}{URL }
\expandafter\ifx\csname urlstyle\endcsname\relax
  \providecommand{\doi}[1]{DOI \discretionary{}{}{}#1}\else
  \providecommand{\doi}{DOI \discretionary{}{}{}\begingroup
  \urlstyle{rm}\Url}\fi
\providecommand{\eprint}[2][]{\urlprefix\url{#1#2}}

\bibitem{fernandez2007epidemiology}
Fernandez, William~G, Yard, Ellen~E and Comstock, R~Dawn.
\newblock \enquote{Epidemiology of lower extremity injuries among US high
  school athletes.}
\newblock \textit{Academic emergency medicine} Vol.~14 No.~7 (2007): pp.
  641--645.

\bibitem{riemann1999examination}
Riemann, Bryan~L, Caggiano, Nancy~A and Lephart, Scott~M.
\newblock \enquote{Examination of a clinical method of assessing postural
  control during a functional performance task.}
\newblock \textit{Journal of Sport Rehabilitation} Vol.~8 No.~3 (1999): pp.
  171--183.

\bibitem{tahmasebi2015evaluation}
Tahmasebi, Razieh, Karimi, Mohammad~Taghi, Satvati, Behnaz and Fatoye, Francis.
\newblock \enquote{Evaluation of standing stability in individuals with
  flatfeet.}
\newblock \textit{Foot \& ankle specialist} Vol.~8 No.~3 (2015): pp. 168--174.

\bibitem{sell2012examination}
Sell, Timothy~C.
\newblock \enquote{An examination, correlation, and comparison of static and
  dynamic measures of postural stability in healthy, physically active adults.}
\newblock \textit{Physical Therapy in Sport} Vol.~13 No.~2 (2012): pp. 80--86.

\end{thebibliography}

%%%%%%%%%%%%%%%%%%%%%%%%%%%%%%%%%%%%%%%%%%%%%%%%%%%%%%%%%%%%%%%%%%%%%%%%%%%%%%%%%%%%%%%

\end{document}